\documentclass[11pt,a4paper]{article}
\usepackage[left=3cm, right=3cm, top=3cm, bottom=3cm]{geometry}
\usepackage{graphicx}
\usepackage{authblk}
\usepackage{float}
\usepackage[T1]{fontenc}
\usepackage[utf8]{inputenc}
\usepackage{amsmath,amsthm,amsfonts,amssymb}
\usepackage{enumerate}
\usepackage{caption}
\usepackage[labelformat=simple]{subcaption}

\title{Closed-form perturbation theory in the Sun-Jupiter restricted three body problem without relegation}\date{}
\author[1]{Irene Cavallari,}
\author[2]{Christos Efthymiopoulos}
\affil[1]{Dipartimento di Matematica, Universit\`a di Pisa}
\affil[2]{Dipartimento di Matematica Tullio Levi Civita, Universit\`a degli studi di Padova}

\def \br{\textbf{r}}
\def\brp{\textbf{r}_P}
\def \bp{\textbf{p}}

\def \aV{a^*}
\begin{document}
		\maketitle
	\begin{abstract}
		We present a closed-form normalization method suitable for the study of the secular dynamics of small bodies inside the trajectory of Jupiter. The method is based on a convenient use of a book-keeping parameter introduced not only in the Lie series organization but also in the Poisson bracket structure employed in all perturbative steps. In particular, we show how the above scheme leads to a redefinition of the remainder of the normal form at every step of the formal solution of the homological equation. An application is given for the semi-analytical representation of the orbits of main belt asteroids. 
	\end{abstract}
	
	\section{Introduction}
	In the present short presentation we summarize our work concerning the development of a normalization method in the framework of the Sun-Jupiter restricted three-body problem (R3BP) in order to represent semi-analytically the secular dynamics of a massless particle inside the planet's trajectory. We search for a normal form not depending on the fast angles of the problem; using modified Delaunay variables, the latter are the mean longitudes of the particle and the planet, $\lambda$ and $\lambda_P$. We will briefly describe below the main steps for the elimination of these angles in the Hamiltonian by a normalization procedure in closed form. A more detailed presentation will be given elsewhere (Cavallari and Efthymiopoulos 2021; in preparation).
	
	In our problem, the initial Hamiltonian has two components, a leading term $Z_0$ not depending on $\lambda$ and $\lambda_P$ and a disturbing function $R$. The dependence of the Hamiltonian on modified Delaunay variables $(\Lambda,\Gamma,\Theta,\lambda,\gamma,\theta)$ and on $\lambda_P$ is through the orbital elements of the particle $(a,e,i,\Omega,\omega,u)$  (with $u$ the eccentric anomaly) and of the planet $(a_P,e_P,f_P)$ (with $f_P$ the planet's true anomaly). We have: 
	\[
	\begin{small}
	Z_0 = -\frac{\mathcal{GM}}{2 a}+n_PI_P, \quad R = \mu \sum_{s=s_0}^{+\infty}\epsilon^sR_s^{(0)}(a,e,i,\Omega,\omega,u,f_P; e_P, a_P), \quad s_0\in\mathbb{Z}^+, \end{small}
	\]
	where $n_P$ is the planet's mean motion and $I_P$ is the dummy action variable conjugated to $\lambda_P$; $\mathcal{M}$ is the mass of the Sun and $\mu=\mathcal{G}m_P$, with ${m}_P$ the mass of Jupiter; $\epsilon$ is a so-called \textit{book-keeping} parameter, i.e. a formal parameter with numerical value $\epsilon=1$ whose powers keep track of the relative size of each perturbing term in $R$. 
	
	\noindent In perturbation theory the normal form is typically computed in an iterative way through a composition of Lie transformations (see \cite{Deprit1969}). At the $j$-th iteration the term $R_{s_0+j-1}^{(j-1)}$ (of book-keeping order $s_0+j-1$) of the Hamiltonian $\mathcal{H}^{(j-1)}$, computed at the previous step, is normalized by means of a Lie generating function $\chi_{s_0+j-1}^{(j)}$, determined by solving the homological equation
	\begin{equation}
	\small
	\mathcal{L}_{\chi_{s_0+j-1}^{(j)}}(Z_0)+\epsilon^{s_0+j-1} R_{s_0+j-1}^{(j-1)}=\epsilon^{s_0+j-1} Z_{s_0+j-1}, 
	\label{homEq0}
	\end{equation}
	with
	\[ \mathcal{L}_{\chi_{s_0+j-1}^{(j)}}(Z_0)=-\Big(n\frac{\partial}{\partial \lambda}+n_P\frac{\partial}{\partial \lambda_P}\Big)\chi_{s_0+j-1}^{(j)},
	\]
	where $n = \sqrt{{\mathcal{GM}}/{a^3}}$ and $\mathcal{L}_{\chi_{s_0+j-1}^{(j)}}=\{\cdot,\chi_{s_0+j-1}^{(j)}\}$ is the Poisson bracket operator. 
	
	\noindent Solving (\ref{homEq0}) in our problem can be complicated since the initial disturbing function $R$ depends on $\lambda$ and $\lambda_P$ through $u$ and $f_P$, which implies solving Kepler's equation in series form to obtain the required trigonometric expansions. Two techniques are typically used to overcome this difficulty. One consists of approximating the original Hamiltonian by means of a Taylor expansion in the eccentricities $e$, $e_P$, to make explicit the trigonometric dependence on the fast angles (see \cite{Brouwer1961}). The drawback of this technique is that it can be used only for low values of the eccentricity. A second technique, introduced in \cite{Palacian1992} and formalized in \cite{Segerman2000, Deprit2001}, is the so-called relegation method: since $n_P<n$ for our problem, the term $n_PI_P$ of the leading term is neglected in equation (\ref{homEq0}), so that this can be solved in closed form. We refer to \cite{Sansottera2017} for a discussion about the algorithm's convergence. 
	
	Here, we propose a closed-form normalization method alternative to relegation, which is suitable for orbits with relatively high values of the eccentricity. A method similar to ours was introduced in \cite{Lara2013}, referring to the motion of a test particle under a multiple expansion of the geopotential (e.g. with $J_2$ and $C_{22}$ terms). In summary, the method works as follows. We perform a multipolar (Legendre) expansion of the initial disturbing function and we expand the semi-major axis as $a=a^*+\delta a$, where $a^*$ is a reference value characteristic of each considered individual trajectory. This last step aims at having constant frequencies in the leading term $Z_0$, which turns out to be useful for algorithmic convenience purposes (see below). The starting Hamiltonian takes the form  
	\begin{equation}
	\small
	\mathcal{H}^{(0)} = Z_0+R, \quad Z_0 =n^*\delta\Lambda+n_PI_P, \quad R=\sum_{s=s_0}^{+\infty}\epsilon^sR_s^{(0)}. \label{H0}
	\end{equation}
	The book-keeping parameter $\epsilon$ separates terms in groups of different order of smallness, depending on four small quantities: $e$, $e_P$, $\delta \Lambda$ and the ratio between the planet and the Sun's masses. To overcome the difficulty of solving (\ref{homEq0}), the main idea is, now, to accept a remainder generated by the homological equation (to be normalized at successive steps):
	\begin{equation}
	\small
	\mathcal{L}_{\chi_{s_0+j-1}^{(j)}}(Z_0)+\epsilon^{s0+j-1}R_{s_0+j-1}^{(j-1)}=\epsilon^{s_0+j-1}Z_{s_0+j-1}+\mathcal{O}(\epsilon^{s_0+j})
	\label{homeq}
	\end{equation}
	where $Z_{s_0+j-1}$ contains the terms of $R_{s_0+j-1}^{(j-1)}$ not depending on $\lambda$ and $\lambda_P$.
	
	The steps to perform  in order to apply the normalization method are described in Section \ref{section: method}. In Section \ref{section: outcomesPCR3BP}, we discuss the applicability of the method in the case of the  planar elliptic R3BP.
	
	\section{Normalization Method}
	\label{section: method}
	\subsection{Hamiltonian preparation}
	We consider a heliocentric inertial reference frame with the $\widehat x$ axis and $\widehat z$ axis parallel to the planet's orbital eccentricity vector and the angular momentum respectively. Let $\br$ be the helicentric position vector, $r=\vert \br \vert$, $\bp$ the vector of the conjugated momentums, $p=\vert \bp \vert$ and $\brp$ the planet's position vector, $r_P=\vert \brp \vert$; the Hamiltonian of the R3BP is
	\[
	\small
	H = \frac{p^2}{2} - \frac{\mathcal{GM}}{r} + \mathcal{R}, \qquad \mbox{with} \quad \mathcal{R}=-\mu\big(\frac{1}{\sqrt{r^2+r_P^2-2\br	\cdot \brp}}-\frac{\br\cdot\brp}{r_P^3}\big)
	\]
	The following operations must be performed:
	\begin{enumerate} 
		\item \textit{Multipolar Expansion:}
		\[
		\small
		\mathcal{R} \sim \mathsf{R} = -\frac{\mu}{r_P}\sum_{j=2}^{o}\frac{r^j}{r_P^j} P_j(\cos \alpha), \quad  \mbox{with} \quad \cos \alpha = \frac{\bf{r}\cdot\bf{r}_P}{rr_P}
		\]
		where $P_j(\cdot)$ are the Legendre functions;
		
		\item \textit{Extended Hamiltonian:}
		the Hamiltonian is expressed as an implicit function of the modified Delaunay variables by means of the orbital elements of the particle and of the planet. A dummy action variable $I_P$, conjugated to $\lambda_P$, is introduced. We get: 
		\[
		\small
		\mathsf{H}=  - \frac{\mathcal{GM}}{2a} +n_PI_P +\mathsf{R}, \quad \mbox{with} \quad \mathsf{R}=\mathsf{R}(a,e,i,\omega,\Omega,u,f_P;a_P,e_P).
		\]
		\item \textit{Expansion of the semi-major axis:}  
		\[
		\small
		\mathcal{H}= n_PI_P + n^*\delta \Lambda -\frac{3}{2}\frac{\delta \Lambda^2}{\aV{^2}} + \dots + \mathsf{R}(\aV + \frac{2\delta \Lambda}{\aV n^*} + \dots, e, i, \omega, \Omega, u, f_P; a_P, e_P), \hspace{1mm} n^*= \sqrt{\frac{\mathcal{GM}}{{\aV}^{3}}}.
		\]
		\item \textit{RM-reduction:} using the identity $r=a(1-e\cos u)$, we re-write $\mathcal{H}$ as
		\[
		\small
		\mathcal{H}= n_PI_P + n^*\delta \Lambda + \Big(-\frac{3}{2}\frac{\delta \Lambda^2}{\aV{^2}} + \dots + \mathsf{R}(\aV + \frac{2\delta \Lambda}{\aV n^*} + \dots, e, i, \omega, \Omega, u, f_P; a_P, e_P)\Big)Q
		\label{H3D}
		\]
		\begin{equation}
		\mbox{with} \quad \small
		Q = \frac{a(1-e\cos u)}{r}=\frac{\aV(1-e\cos u)}{r}+2\frac{(1-e\cos u)}{\aV n^* r}\delta \Lambda + ...=1.
		\label{Qdef}
		\end{equation} 
	\end{enumerate}
	\subsection{Book Keeping}
	To write the initial Hamiltonian in the form~(\ref{H0}), we use the book-keeping parameter $\epsilon$ to keep track of the relative size of the several terms of $\mathcal{H}$. There are four different small parameters to consider in the problem: $\mu$, $\delta\Lambda$, $e$ and $e_P$. We define
	$
	s_0 =\Bigg\lceil \frac{\log\big(\frac{{m}_P}{\mathcal{M}}\big)}{\log (e)}\Bigg\rceil
	$
	and we adopt the following book-keeping rules: 
	\begin{itemize}
		\item terms depending on $e^je_P^k$ ($j,k\in\mathbb{Z}$) are multiplied by $\epsilon^{(j+k)}$;
		\item terms depending on $(1+\eta)^j$ and $(1-\eta)^k$, with $\eta=\sqrt{1-e^2}$, ($j,k\in\mathbb{N}$) are multiplied by $\epsilon^0$ and $\epsilon^{2k}$ respectively; 
		\item terms depending on $\mu^j\delta\Lambda^k$ ($j,k\in\mathbb{N}$) are multiplied by $\epsilon^{(j+k)s_0}$;
		\item terms depending on $\delta \Lambda^k$ ($k\in\mathbb{N}$) are multiplied by $\epsilon^{(k-1)s_0}$.
	\end{itemize} 
	Finally, the value of $s_0$ is specified, for any particular trajectory, by the initial value $e(0)$.
	
	\subsection{Poisson bracket structure}
	\label{sectionPB}
	During the normalization process, we need to compute Poisson brackets of the form $\{A_1,A_2\}$, where $A_1$ and $A_2$ are implicit functions of $(\delta\Lambda,\Gamma,\Theta,I_P,\lambda,\gamma,\theta,\lambda_P)$ through the variables $(e,i,\omega,\Omega,u,f_P)$, $r$, $\eta=\sqrt{1-e^2}$, $\phi=u-M$ (with $M$ the mean anomaly). 
	To compute $\{A_1,A_2\}$, we use the formula:
	\begin{equation}
	\begin{split}
	\{A_1,A_2\}= & \frac{\partial A_1}{\partial \lambda}\frac{\partial A_2}{\partial \delta \Lambda}-\frac{\partial A}{\partial \delta \Lambda}\frac{\partial A_2}{\partial \lambda}+\frac{\partial A_1}{\partial \gamma}\frac{\partial A_2}{\partial \Gamma}-\frac{\partial A_1}{\partial \Gamma}\frac{\partial A_2}{\partial \gamma}+\frac{\partial A_1}{\partial \theta}\frac{\partial A_2}{\partial \Theta}-\frac{\partial A_1}{\partial \Theta}\frac{\partial A_2}{\partial \theta}\\
	& +\Big(\frac{\partial A_1}{\partial \lambda_P}\frac{\partial A_2}{\partial I_P}-\frac{\partial A_1}{\partial I_P}\frac{\partial A_2}{\partial \lambda_P}\Big)Q.
	\end{split}
	\label{poissbracket}
	\end{equation}
	To evaluate the partial derivates with respect to the modified Delaunay variables in the formula, we must perform a composition of partial derivates. The last term is multiplied by Q defined in (\ref{Qdef}) to allow significant simplifications to be carried out automatically during the normalization process. For the same reason, the following relations must be used for any $A=A_{1,2}$ in the computation of the partial derivatives: 
	\begin{small}
		\[
		\frac{\partial e}{\partial \delta \Lambda}= -\frac{\eta e}{(1+\eta)n^*{\aV}^2}\epsilon +\mathcal{O}(\epsilon^{s_0}\delta \Lambda),  \hspace{1mm} \frac{\partial e}{\partial \Gamma}=-\frac{\eta}{a^{*^2} n^* e}	\epsilon^{-1} +\mathcal{O}(\epsilon^{s_0}\delta\Lambda),  
		\hspace{1mm} \frac{\partial \eta}{\partial \delta \Lambda}= \frac{1-\eta}{n^*{\aV}^2}\epsilon^2+\mathcal{O}(\epsilon^{s_0}\delta\Lambda),
		\]
		\[
		\frac{\partial \eta}{\partial \Gamma}=\frac{1}{a^{*^2} n^*}+\mathcal{O}(\epsilon^{s_0}\delta\Lambda), \hspace{1mm}
		\frac{\partial \cos i}{\partial \delta \Lambda}=\frac{1-\cos i}{a^{*^2} n^* \eta} +\mathcal{O}(\epsilon^{s_0}\delta), \hspace{1mm}
		\frac{\partial \cos i}{\partial \Gamma}=\frac{\cos i-1}{a^{*^2} n^* \eta} +\mathcal{O}(\epsilon^{s_0}\delta\Lambda), 
		\]
		\[ \frac{\partial  \cos i}{\partial \Theta}=-\frac{1}{a^{*^2} n^* \eta} +\mathcal{O}(\epsilon^{s_0}\delta\Lambda),\hspace{1mm}
		\frac{\partial u}{\partial 	\lambda}=\frac{a^*}{r}+\mathcal{O}(\epsilon^{s_0}\delta\Lambda), \hspace{1mm} \frac{\partial u}{\partial 	\gamma}=\frac{a^*}{r}+\mathcal{O}(\epsilon^{s_0}\delta\Lambda),\hspace{1mm} \frac{\partial \phi}{\partial \lambda}=\frac{\aV}{r}-1+\mathcal{O}(\epsilon^{s_0}\delta\Lambda),
		\]
		\[\frac{\partial \phi}{\partial \gamma}= \frac{a^*e\cos u}{r}\epsilon+\mathcal{O}(\epsilon^{s_0}\delta\Lambda),\hspace{1mm} \frac{\partial u}{\partial \delta \Lambda}=\frac{\eta e\sin u}{r(1+\eta)n^*{\aV}}\epsilon+\mathcal{O}(\epsilon^{s_0}\delta\Lambda),
		\hspace{1mm} \frac{\partial u}{\partial \Gamma}=\frac{\eta\sin(u)}{a^*n^*er}\epsilon^{-1}+\mathcal{O}(\epsilon^{s_0}\delta\Lambda), 
		\]
		\[ \frac{\partial \phi}{\partial \delta \Lambda}=\frac{\eta e\sin u}{r(1+\eta)n^*{\aV}}\epsilon+\mathcal{O}(\epsilon^{s_0}\delta\Lambda), \hspace{1mm} \frac{\partial \phi}{\partial \Gamma}=\frac{\eta\sin(u)}{a^*n^*er}\epsilon^{-1}+\mathcal{O}(\epsilon^{s_0}\delta\Lambda), \hspace{1mm} \frac{\partial \phi}{\partial \delta \Lambda}=\frac{\eta e\sin u}{r(1+\eta)n^*{\aV}}\epsilon+\mathcal{O}(\epsilon^{s_0}\delta\Lambda),
		\]
		\[ \frac{\partial \phi}{\partial \Gamma}=\frac{\eta\sin(u)}{a^*n^*er}\epsilon^{-1}+\mathcal{O}(\epsilon^{s_0}\delta\Lambda), \hspace{1 mm}
		\frac{\partial r}{\partial \lambda}=\frac{{\aV}^2e\sin(u)}{r}\epsilon +\mathcal{O}(\epsilon^{s_0}\delta\Lambda), \hspace{1 mm} \frac{\partial r}{\partial \gamma}=\frac{{\aV}^2e\sin(u)}{r}\epsilon+\mathcal{O}(\epsilon^{s_0}\delta\Lambda),\]
		\[
		\frac{\partial r}{\partial \delta \Lambda}= \frac{\eta e\cos u}{(1+\eta)n^*{\aV}}\epsilon+\mathcal{O}(\epsilon^{s_0}\delta\Lambda), \hspace{1mm} \frac{\partial r}{\partial \Gamma}=\frac{\eta(e\epsilon-cos(u))}{n^*er}\epsilon^{-1}+\mathcal{O}(\epsilon^{s_0}\delta\Lambda),\frac{\partial \omega}{\partial \gamma} = -1, \hspace{1mm} \frac{\partial \omega}{\partial \theta} = 1,
		\]
		\[
		\frac{\partial \Omega}{\partial \theta} = -1, \hspace{1 mm}
		\frac{\partial f_P}{\partial \lambda_P}=1+\frac{2e_P\cos(f_P)}{\eta_P^3}\epsilon+\Bigg(\frac{1}{\eta_P^3}-1+\frac{e_P^2\cos^2(f_P)}{\eta_P^3}\Bigg)\epsilon^2.
		\]
	\end{small}
	The fact that the book-keeping parameter $\epsilon$ is present in all partial derivatives of terms depending on $e$, $e_P$ and $\delta \Lambda$ is an essential element of the method. In fact, we can readily see that, for every case in which $s_0>1$, the result of Poisson brackets between such terms contains terms of different powers of $\epsilon$, which, however, are always larger than the current normalization order (for the case $s_0=1$, instead, see Cavallari and Efthymiopoulos, 2021; in preparation).
	
	\subsection{Normalization Process}
	At the ${j}\mbox{-th}$ iteration of the normalization process, we must determine the generating function $\chi_{s_0+j-1}^{(j)}$ satisfying the homological equation (\ref{homeq}). The remainder term ${R}_{s_0+j-1}^{(j-1)}$ contains four different types of terms:
	\begin{itemize}
		\item type 1:  $\frac{\aV}{r}f(e,i,\eta,\omega,\Omega)$,
		\item type 2:  $ \frac{\aV}{r} \widehat{f}_{\bf{k}}(e,i,\eta)\cos(k_1u+k_2f_P+k_3\omega+k_4\Omega)$,
		\item type 3:  $\frac{\aV}{r^p}\bar{f}(e,i,\eta,\omega,\Omega)$, $p>1$,
		\item type 4:  $ \frac{\aV}{r^p} \tilde{f}_{\bf{k}}(e,i,\eta)\cos(k_1u+k_2f_P+k_3\omega+k_4\Omega)$, $p>1$.
	\end{itemize} 
	The corresponding terms added in $\chi_{s_0+j-1}^{(j)}$ are
	\begin{itemize}
		\item for type 1: $\frac{1}{n^*}f(e,i,\eta,\omega,\Omega)\phi$,
		\item for type 2:  $ \frac{1}{k_1n^*+k_2n_P} \widehat{f}_{\bf{k}}(e,i,\eta)\cos(k_1u+k_2f_P+k_3\omega+k_4\Omega)$,
		\item for type 3:  $\frac{1}{n^*}\phi\sum_{k=1}^{p}\frac{\hat{f}(e,\eta,i,\omega,\Omega)}{a^{k-1}r^{p-k}}$, $p>1$,
		\item for type 4:  $\frac{1}{k_1n^*+k_2n_P}\frac{1}{{r}^{p-1}} {\tilde{f}_{\bf{k}}(e,\eta,i)\sin(k_1u+k_2f_P+k_3\omega+k_4\Omega)}$, $p>1$.
	\end{itemize} 
	The new Hamiltonian is 
	\[
	\small
	\mathcal{H}^{(j)}=exp(\mathcal{L}_{\chi_{s_0+j-1}^{(j)}})\mathcal{H}^{(j-1)}=Z_0+\sum_{s=s_0}^{s_0+j-1}\epsilon^sZ_{s}+\sum_{s=s_0+j}^{r}\epsilon^s{{R}}_{(s)}^{(j)}.
	\]
	where $\exp(\mathcal{L}_{\chi})$ denotes the operation
	$
	\exp(\mathcal{L}_{\chi})=\sum_{k=0}^{\infty}\frac{1}{k!}\mathcal{L}_{\chi}^k$, with $ \mathcal{L}_{\chi}^k = \underbrace{\{\dots\{\{\cdot,\chi\},\chi\}\dots,\chi\}}_{k\hspace{0.5mm}\mbox{times}}$.
	
	\begin{figure}[h!]
		\centering
		\begin{subfigure}{0.3\textwidth}
			\centering
			\begin{minipage}{\textwidth}
				\includegraphics[width=\textwidth]{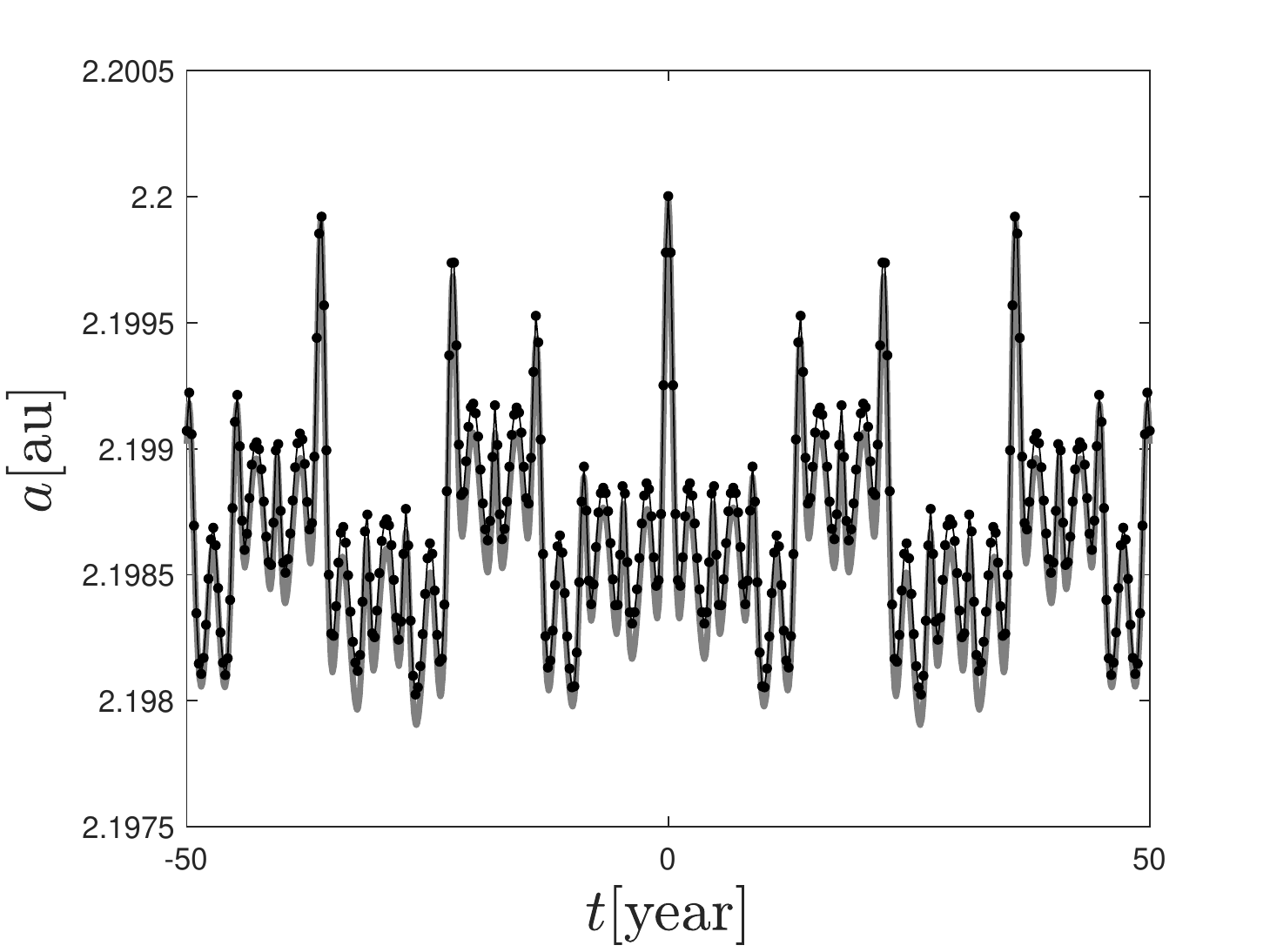}
			\end{minipage}
			\begin{minipage}{\textwidth}
				\includegraphics[width=\textwidth]{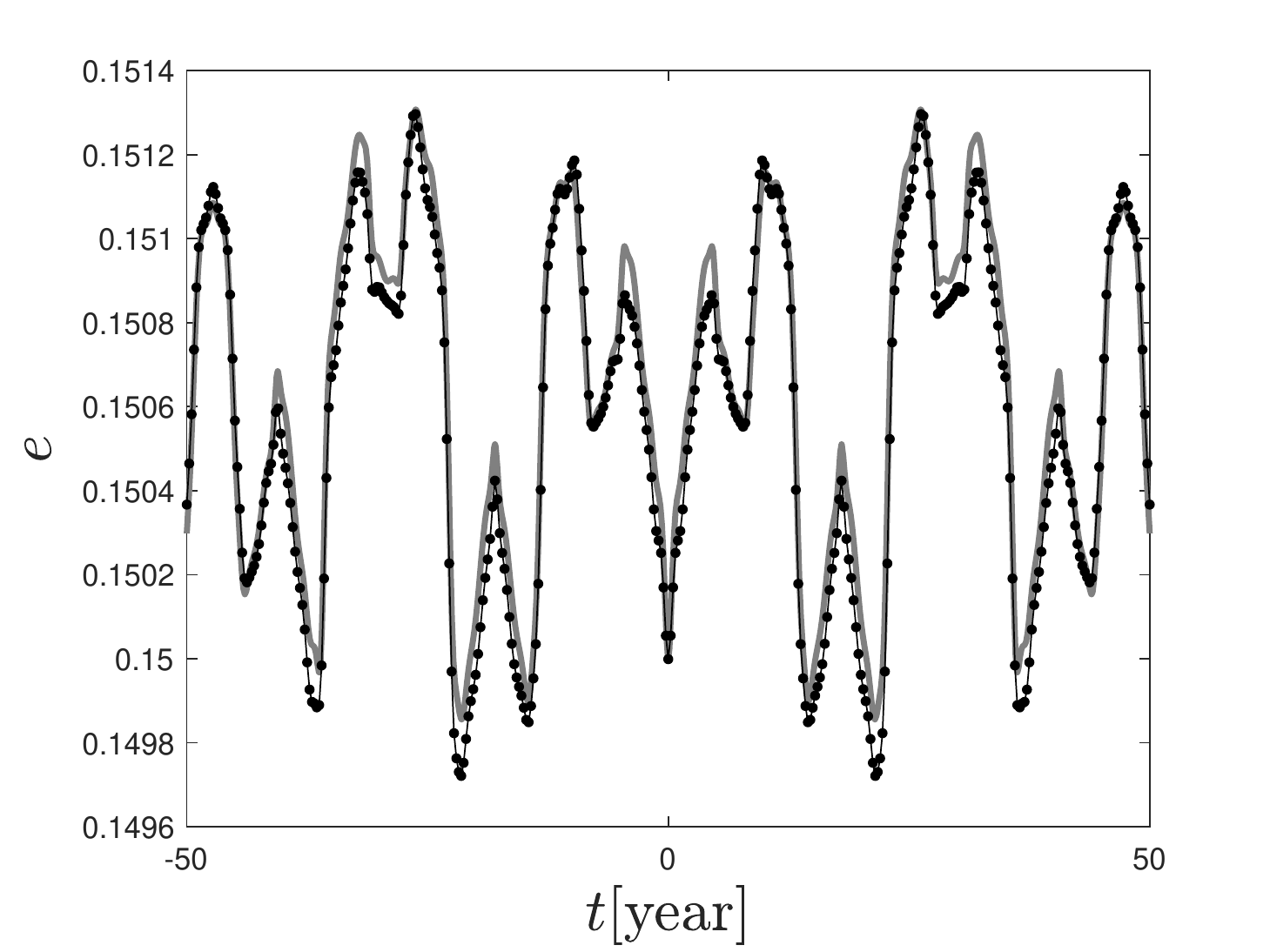}
			\end{minipage}
			\subcaption*{ case 1}
		\end{subfigure}
		\begin{subfigure}{0.3\textwidth}
			\begin{minipage}{\textwidth}
				\includegraphics[width=\textwidth]{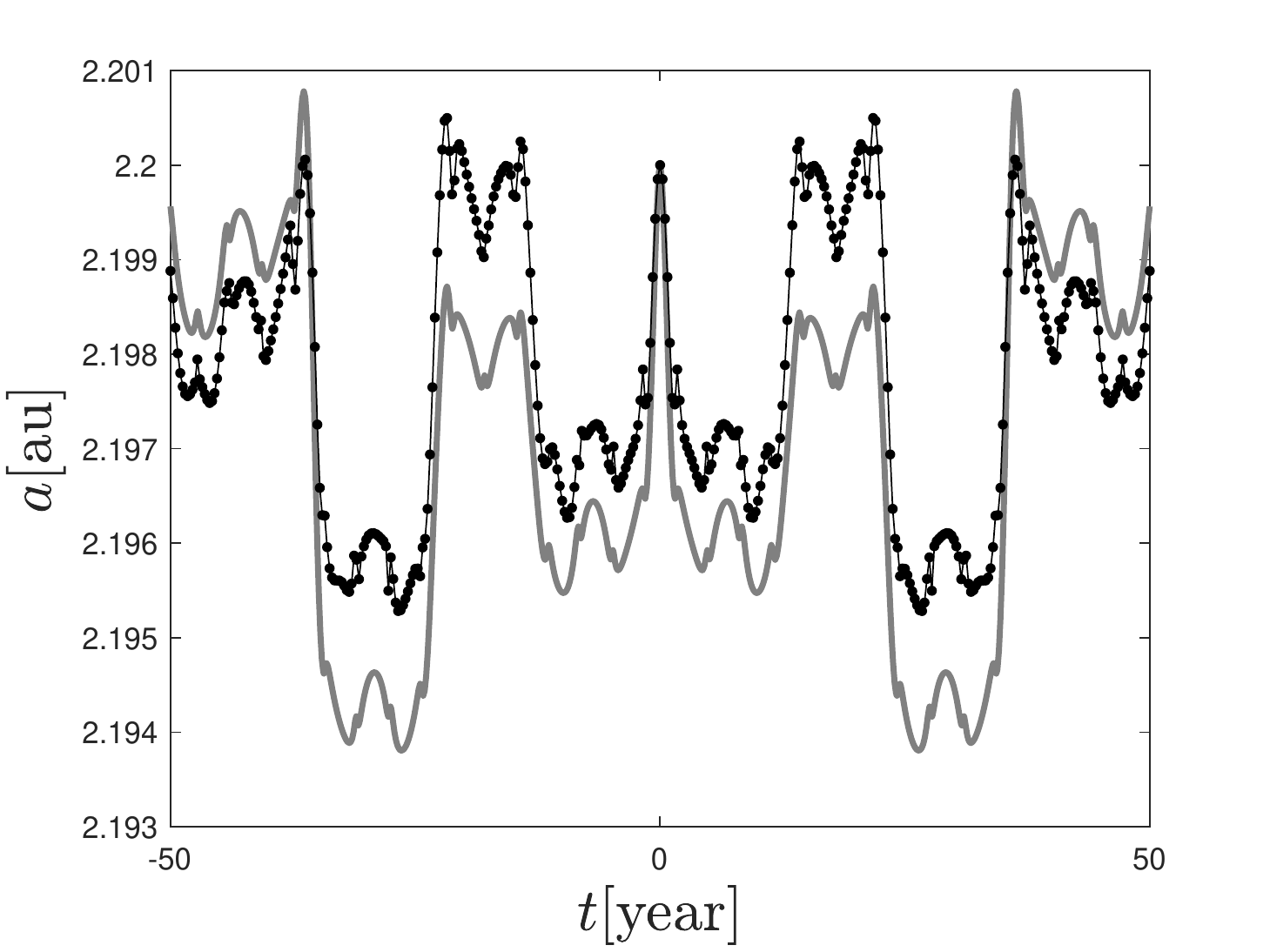}
			\end{minipage}
			\begin{minipage}{\textwidth}
				\includegraphics[width=\textwidth]{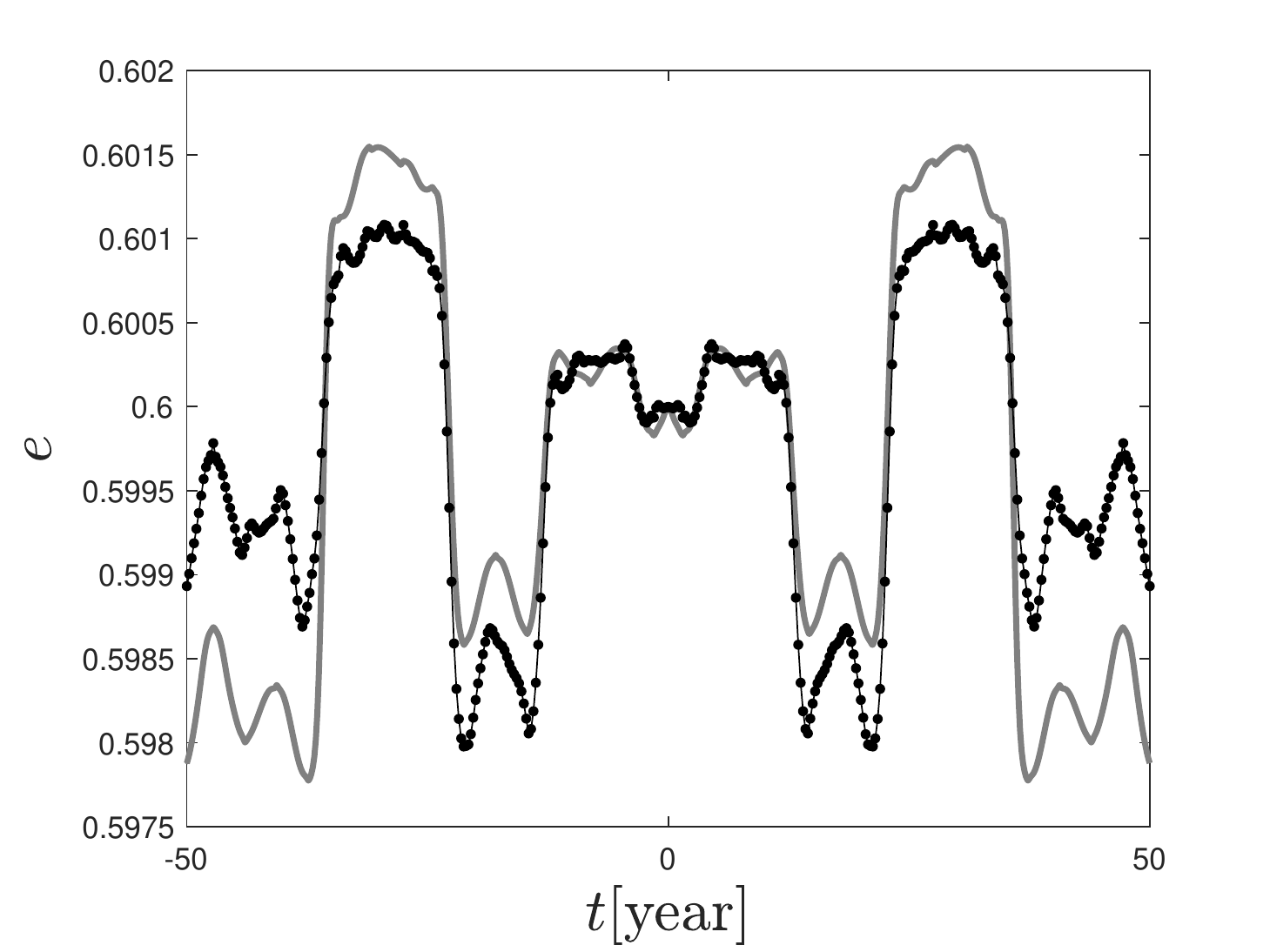}
			\end{minipage}
			\subcaption*{ case 2}
		\end{subfigure}
		\begin{subfigure}{0.3\textwidth}
			\begin{minipage}{\textwidth}
				\includegraphics[width=\textwidth]{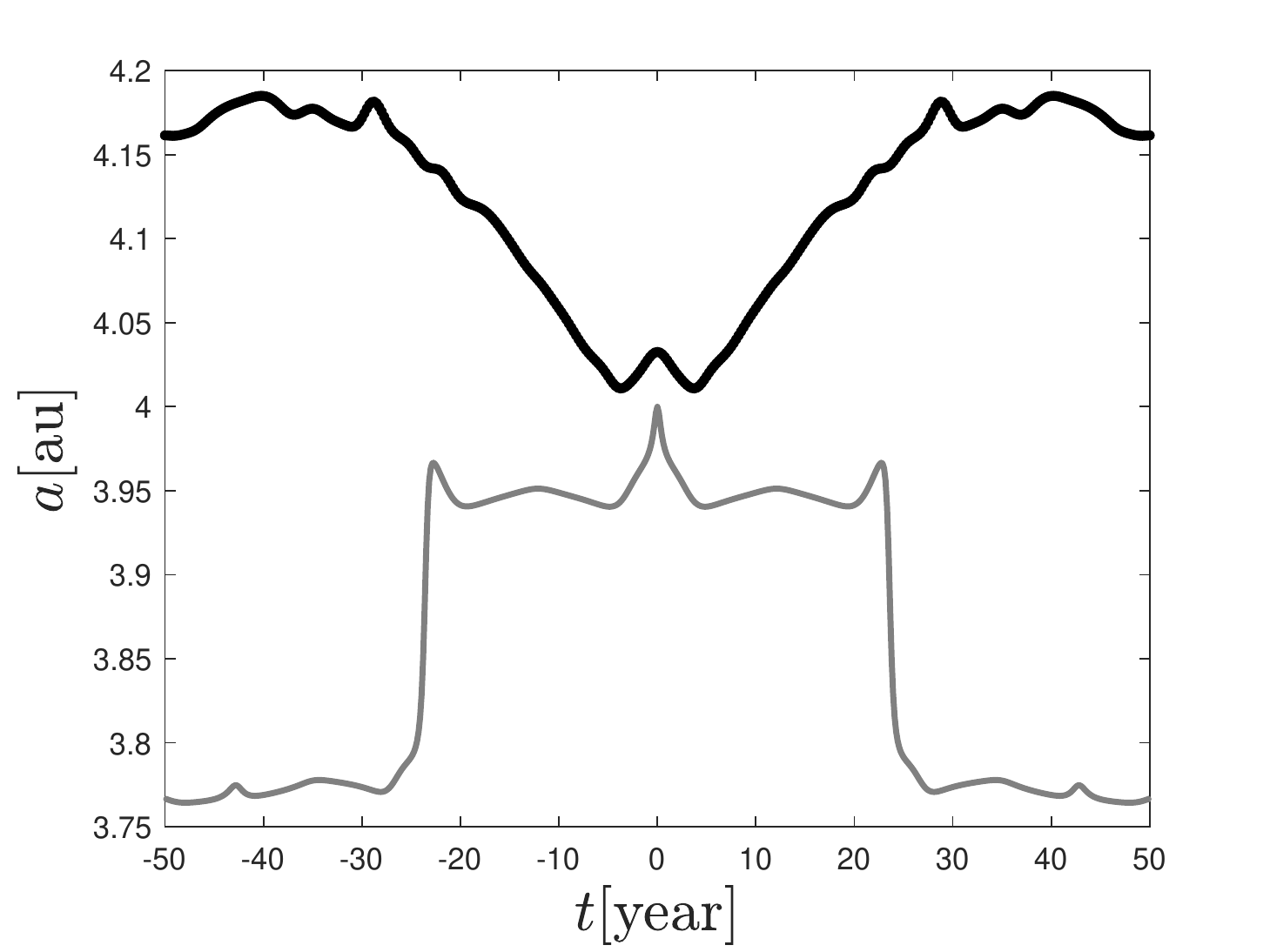}
			\end{minipage}
			\begin{minipage}{\textwidth}
				\includegraphics[width=\textwidth]{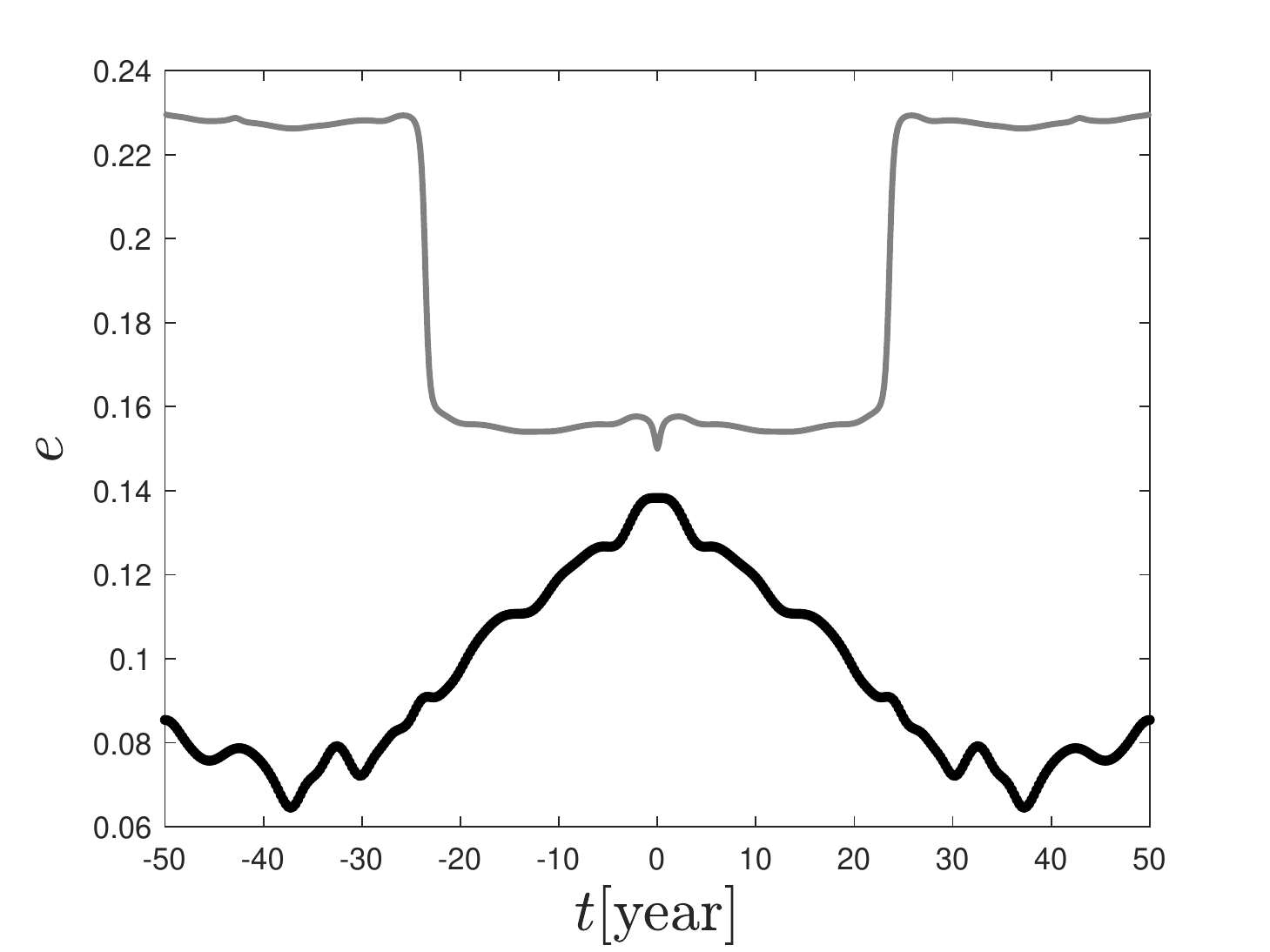}
			\end{minipage}
			\subcaption*{ case 3}
		\end{subfigure}
		\caption{Evolution of the semi-major axis and of the eccentricity 
			for orbits with the following initial conditions:  (case 1) $a(0)=2.2$au, $e(0)=0.15$; (case 2) $a(0)=2.2$au, $e(0)=0.5$; (case 3) $a(0)=4$au, $e(0)=0.15$. For all of them we set $a^*=a(0)$ and $\omega(0) = 90^{\circ}$, $M(0)=90^{\circ}$, $i(0)=20^{\circ}$, $\Omega(0)=0^{\circ}$, $\lambda_P(0)=0^{\circ}$. We show the evolution computed through a numerical propagation of the trajectory (grey line) and semi-analytically through the normal form computed through the algorithm of section \ref{section: method} (black line). }
		\label{fig: r3bp_irene}
	\end{figure}
	
	\section{Results}
	\label{section: outcomesPCR3BP}
	The error of the method increases with an increase of the semi-major axis or the eccentricity. The trend with respect to the semi-major axis has two main causes: first of all, increasing $a$, we obtain trajectories which arrive closer to the planet and to the region of Hill-unstable orbits. The other reason is related to the multipolar expansion: as $a$ increases, the heliocentric distances of the particle and of the planet become comparable. Dealing with this problem requires performing a multipolar expansion of high order, which, however, implies a substantial increase in the number of terms to normalize. Similarly, reducing the error for highly eccentric orbits requires performing a large number of normalization steps. However, the consequent increase of the computational time can become significant. 
	In the case of the Sun-Jupiter planar elliptic R3BP, performing a multipolar expansion of order equal to $5$ and doing from $4$ to $7$ normalization steps allows to obtain accurate results up to an initial semi-major axis $a\le
	0.6a_P$ and for relatively high values of $e$, up to almost $0.7$. 
	
	\noindent Figure \ref{fig: r3bp_irene} shows the evolution of the semi-major axis and of the eccentricity in three cases: a trajectory with a low value of the eccentricity, one with a relatively high value of eccentricity and one with a semi-major axis larger than $0.6a_P$. We compare the outcomes obtained semi-analytically through our normal form algorithm with those resulting from a numerical propagation. The method works well in the first two cases, while the errors increase in general with $e$. In fact, in case 1 the maximum relative error is $\sim 10^{-4.6}$ for the semi-major axis and $\sim 10^{-3.34}$ for the eccentricity, while in case 2 it is $\sim 10^{-3.5}$ for the semi-major axis and $\sim 10^{-2.75}$ for the eccentricity. Instead, in case 3 the method is not sufficiently precise: the maximum relative error is $\sim 10^{-0.97}$ and $\sim 10^{-0.2}$ for the semi-major axis and for the eccentricity respectively. More detailed examples and results are contained in (Cavallari and Efthymiopoulos, 2021; in preparation).

	\section*{Acknowledgements}
	I. Cavallari has been supported by the MSCA-ITN Stardust-R,
	Grant Agreement n. 813644 under the H2020 research and innovation
	program. C. Efthymiopoulos acknowledges the support of MIUR-PRIN 20178CJA2B "New frontiers of Celestial Mechanics: theory and applications".

\end{document}